\documentstyle[12pt]{article}

	 \def\be{\begin{equation}}
	 \def\bea{\begin{eqnarray}}
	 
	 \def\o{\over}
	 \def\ee{\epsilon}

	 \def\ee{\end{equation}}
	 \def\eea{\end{eqnarray}}
	 \def\R{\rm {I\kern-.200em R}}
	 \def\C{\rm {I\kern-.520em C}}
\begin{document}
\pagestyle{empty}
\begin{flushright} 
{BROWN-HET-1025} \\
{IPM-95-116}\\
{June 1996}  \\
\end{flushright} 
\vspace*{5mm}
 \begin{center} 
{\bf  Remarks on a Decrumpling Model of the Universe}
\\ [10mm] 
\renewcommand{\thefootnote}{\alph{footnote}} 
J.A.S. Lima$^{1,2,}$\footnote{e-mail:limajas@het.brown.edu},
 M.
Mohazzab$^{1,3,}$\footnote{e-mail: mohazzab@het.brown.edu}
\end{center}

\begin{flushleft} 

1){\it Physics Department, Brown
University,Providence RI. 02912. USA.} \\
2){\it Departamento de Fisica,
Universidade Federal do Rio Grande do Norte, 59072-970, C.P. 1641 Natal,
Rio Grande do Norte, Brazil.}\\ 
3){\it
  { Institute for Studies in Theoretical Physics and
	      Mathematics,
	     }{
	     P.O.Box  5746, Tehran 19395, Iran.
	     }{
		
\&}{  Physics Dept. Alzahra University 
	      }{              
	       Tehran 19834, Iran}}

\end{flushleft} 
\begin{center} 
{\bf 
Abstract}
\end{center} 
\vspace*{3mm}

It is argued that when the dimension of space is a constant integer
the full set of Einstein's field equations has more information than 
the spatial components of Einstein's equation plus the 
energy conservation law. Applying the former approach to decrumpling FRW cosmology recently proposed, it is 
shown that the spacetime singularity cannot be avoided and that 
turning points are absent. This result is in contrast to the decrumpling nonsingular spacetime model with turning points previously obtained using the latter approach.

\setlength{\textheight}{7.60in}
\newpage\setcounter{page}{1}\pagestyle{plain}
\renewcommand{\thefootnote}{\arabic{footnote}}

\section{ Introduction}

In the standard model of cosmology, the spacetime is usually assumed to be a 
differentiable manifold even at the very early stages of the evolution 
of the universe. However, many studies on the underlying structure of the
spacetime suggest a more complex nature e.g., foam like\cite{QS} or even a fractal structure\cite{FS,CP}. 

Recently, a cosmological model has
been proposed assuming that the basic  building blocks of the spacetime has 
a cellular structure\cite{KMME}. In the early universe this cellular space is crumpled so that a fractal (and therefore non-integer) dimension  
can  naturally be defined for the whole space.  In such a work the dimension of the space 
starts from a large (but not infinity) number when the 
 universe is a minimum in its size.  The expansion of the universe causes the 
 wrinkled space to decrumple while the dimension of space decreases very fast
to the observational value.

The Lagrangian equations of the model lead to an oscillatory universe (a universe with turning points) which may solve the horizon and the big bang singularity  problems. Later on, this scenario was extended to the class of multidimentional cosmological models \cite{BMR} where extra factor spaces play the 
role of matter fields.  In this multidimensional cosmological model, an
inflationary solution was found together with the prediction that the 
universe starts from  a nonsingular spacetime.

In the approach of reference \cite{KMME} the leading equations are a generalization of the 
space components of the Einstein field equations(EFE) to $D$ variable 
dimensions plus the generalized equation expressing the energy 
conservation law for a self-gravitating fluid.

In this letter we argue that the use of the space 
components of EFE plus the energy conservation law lead to results with
less information than the use of the full set of equations. As 
we known, the time(00) component of the EFE is a constraint which can be 
easily implemented in the most simple cases, as happens for the class of Friedmann-Robertson-Walker(FRW) spacetimes. However, for more complex 
situations, like in the case 
of a decrumpling universe which is endowed with a fractal structure, such approaches give rise to completely different physical properties. At first sight, one may argue that the discrepancy may arise from the fact 
that the generalization of the EFE to noninteger 
dimensions is not uniquely defined. However, we believe that the puzzle may
in principle be solved if the constraint equation is further implemented when the energy 
conservation law has ab initio been considered.

In what follows, we first make explicit the difference between the two schemes 
in the case of the standard FRW model, and then 
an extension to the case of a decrumpling universe it will be discussed.

\section{ The Problem}

For the sake of simplicity, let us consider the 3-dim 
spatially flat FRW line element:

\be \label{E1}
ds^2= dt^2 - a^2(t)(dx^{2} + dy^{2} + dz^{2}) \quad ,
\ee
where $a(t)$ is the scale factor. In this background, the non-trivial 
EFE for a comoving perfect fluid and the 
energy conservation law ($u_{\alpha}{T^{\alpha\beta}}_{{;}\beta}=0$), 
which is contained in the 
EFE, may be 
written as (in our units $8 \pi G=c=1$)

\be \label {E2}
\rho = 3(\frac{\dot{a}}{a})^{2} \quad,
\ee

\be \label{E3}
p = -2\frac{\ddot{a}}{a} - (\frac{\dot{a}}{a})^{2}
\quad,
\ee

\be  \label{E4}
\dot\rho + 3(\rho +p)\frac{\dot a}{a} = 0  \quad,
\ee
where an overdot means time derivative and $\rho$ and $p$ are 
the energy density and pressure, respectively. From Bianchi identities, we
known that (2.4) can be obtained of (2.2) and (2.3), by just eliminating the second derivative of the scale factor.
As usual, it will be assumed that the material medium obeys
the  barotropic equation of state 
 
\be \label{ES} 
p=(\gamma - 1) \rho  \quad,
\ee
where $0 \leq \gamma \leq 2$ is the ``adiabatic index''. 
Combining the above equation with the EFE 
 (\ref{E2}) and (\ref {E3}), we get the FRW differential equation
describing the evolution of the scale factor
\be \label{OK}
a \ddot a +({{3\gamma-2 \o 2}})\dot a^2 =0  \quad.
\ee

On the other hand, using (\ref{ES}), the energy conservation law (\ref{E4}) 
yields for the energy density $\rho={B\over a^{3\gamma}}$ and 
for the pressure $p=(\gamma-1){B\over a^{3\gamma}}$, 
where $B$ must be positive due to the weak energy condition($\rho>0$). Now, inserting this value of 
$p$ into (\ref{E3}) one obtains a quite different 
equation of motion 

\be \label{W}
a \ddot a + \frac{\dot a^2}{2} + \frac{{(\gamma-1) B a^{2-3\gamma}}}{2}=0 
\quad.
\ee
which reduces to (\ref{OK}) only in the case of
a dust-filled universe($\gamma$=1).

As one may check, the above equations (\ref{OK}) and (\ref{W}) have, respectively, the following first integrals:

\be \label{O1}
\dot a^2={A a^{2-3\gamma}} \quad, 
\ee
and
\be \label{W1}
\dot a^2={F\o a} + \frac {B a^{2-3\gamma}}{3}  \quad,
\ee
where $A$ and $F$ are two $\gamma$-dependent constants.
As expected, if $\gamma=1$  equations 
(\ref{O1}) and (\ref{W1}) become identical up to trivial identifications. 
Note also from (\ref{E2}) that $A$ (like $B$) is positive definite, whereas in the approach 
leading to (\ref{W1}) the sign of 
$F$ is arbitrary. If $F$ is negative, for instance, (\ref{W1}) has a turning point i.e.
a specific value of the scale 
factor, say $a^{*}$, for which $\dot a(a^*)=0$ and the motion is inverted. How this fictitious turning point is avoided? Only if one uses the constraint given by the $(00)$ component of the EFE. As easily seen, such ambiguity is
completely solved by using (\ref{E2}) since it 
implies that $F=0$ with (\ref{W1}) reducing to (\ref{O1}) for $B=3A$. It thus follows  that 
only the full set
of EFE fix the unique physical solution for
a given problem. Hence, if one uses the energy conservation law in a 
non trivial situation e.g., for anisotropic or inhomogeneous models, the constraint equation need to be further satisfied. Although
quite familiar for cosmologists working with exact solutions, this result is apparently not well known, and potentially, it  may generate 
paradoxes when new ingredients are added to the FRW geometries. This point will now be exemplified for the case of a decrumpling universe.

\section {New setup of the equations of motion}

In \cite{KMME}, using the space component of Einstein equation together 
with the energy conservation equation yields an oscillatory universe.  
Unlike the result of the previous section, the reason for finding two 
turning points is not the lack of knowledge (or not using) 
the time component of the Einstein equation, 
because the existence of the turning point does not depend on any constant. 

Here we would like to change the approach of \cite{KMME}, by considering 
the time component of the Einstein equation instead of the energy 
conservation equation.

Let us now consider the D-dimensional FRW flat line element:
\be
ds^2= dt^2 - a^2(t)\delta_{ij}dx^{i}dx^{j} \quad,
\ee
where $i,j=1,2,...D.$

The generalized Lagrangian is given by \cite{KMME}

\be
{L\over{a_0^{D_0}}}=-{{D(D-1)}\over{2\kappa}}\Big({{\dot a}\over a}\Big)^2
   \Big({a\over{a_0}}\Big)^D+\Big( -{{\hat\rho}\over 2}+{{\hat p D a^2}\over
   2}\Big)=:{\cal L} \quad,
\ee

where
\be
\hat\rho:=\rho\Big({a\over{a_0}}\Big)^D \quad,
\ee
and
\be
    \hat p:=p a^{-2}\Big({a\over{a_0}}\Big)^D \quad,
\ee
with the constraint
\be
\Big({a\over\delta}\Big)^D=e^C \quad.
\ee
In the above equations, $a_0$ and $D_0$ are the present 
values for the scale factor and the dimension of the universe, respectivley, $\delta$ is a fundamental length and $C$ is a dimensionless constant, which could be determined from observations. As shown in the Ref.\cite{KMME},  in the 
limit $C\rightarrow \infty$ the standard 
D-dimensional FRW model is recovered.

{F}rom (3.11) one obtains the following 
equation of motion:
\be \label{ii}
(D-1)\bigg\{ {{\ddot a}\over a}+\Big[ {{D^2}\over{2D_0}}-1-
  {{D(2D-1)}\over{2C(D-1)}}\Big]\Big({{\dot a}\over a}\Big)^2\bigg\} +
  \kappa p\Big( 1-{D\over{2C}}\Big) =0 \quad,
\ee
which is equivalent to the spacelike components
of the Einstein equations. 
Naturally, (3.15) is also the analogous to (\ref{E3}) in the 3-dimensional 
formulation. As remarked earlier, at this 
point we do not insert  the expression 
for the pressure obtained by combining  the 
energy conservation law with 
the equation of state as 
has been done in Ref.\cite{KMME}. In order to obtain the alternative 
differential equation we
notice that the (00)-component of the Einstein equation can be written as
($D \neq 1$)
\be \label{00}
 \Big({{\dot a}\over a}\Big)^2={{2\kappa\rho}\over{D(D-1)}} \quad.
\ee
To get (\ref{00}), we may simply variate the Lagrangian with respect to a 
lapse function.  It is clear that this lapse function does not interfere 
either with the variation with respect to the scale factor 
or with the dimension.

Therefore, the equation (\ref{00}) is the same 
generalization of the (00)-component of the EFE in $D$ dimensions.
We stress that in the case of integer dimension, the
above equations ((\ref{ii})
and (\ref{00})) also contain the energy 
conservation law. 

Now, combining (3.15), (3.16) and the constraint (2.5) we get the following 
evolution equation for the scale factor

\be
{\ddot a \o a}+ ({D^2\o {2D_0}}-1  -{{D(2D-1)}\o {2C(D-1)}} 
+{1\over 2}(\gamma_{(D)} -1)D(1-{D\over {2C}}))({\dot a \o a})^2=0 \quad,
\ee
where $\gamma_{(D)}$  is the ``adiabatic index'' for D spatial dimensions.
It is clear that in the limit $C\rightarrow  \infty$, we recover the FRW equation (\ref{OK}).
The above equation should be compared with (II-6) of Ref.\cite{KMME}.

The first integral of (3.17) is

\be \label{TU}
({\dot a \o a})^2 = E e^{2 C \int dD {1+f(D)\o {D^2}}} \quad,
\ee
where the function $f(D)$ reads
\be
f(D)={D^2\o {2D_0}}- 1 -{{D(2D-1)}\o {2C(D-1)}}
+{1 \o 2}(\gamma_{(D)}-1)D(1-{D\o {2C}}) \quad.
\ee

Inserting (3.19) into (3.18) and assuming a 
radiation dominated universe 
(i.e. $\gamma_{(D)}=1+{1\over D}$), we find 

\be \label{Tu1}
({\dot a \o a})^2 = E {e^{{CD\o {D_o}}-{C\o D}} 
\o D^2(D-1) } \quad,
\ee
and combining this result with 
the constraint expression (3.14) it follows that 
\be
\dot D^2=D^2 E  {e^{{CD\o {D_o}}-{C\o  D}} 
\o {C^2(D-1)} } \quad.
\ee
The above equations should be compared with the equivalent ones of Ref.\cite{KMME}, which were obtained  using the energy conservation law. As one may check, in the FRW limit, equation (3.21) yields $\dot D=0$, whereas (3.20) reduces to (2.8) as it should be expected. Here as there, the nonexistence of turning points may easily be determined by examining the 
qualitative behavior of the above coupled first integrals. In this connection, we remark that (3.21) may be interpreted as the energy equation of a point
mass system having one degree of freedom
whose potential energy is given by 
 
\be \label{pot}
V(D)=-D^2 {e^{{CD\o {D_o}}-{C\o  D}} 
\o {2C^2(D-1)} } \quad,
\ee
and the total energy $H=\frac{1}{2}  \dot D^2 + V(D)$
is identically zero. This potential (\ref {pot}) is singular at $D=1$ 
and tends to $-\infty$ either if $D \rightarrow 1$ or $D \rightarrow \infty$. Since it is always 
negative there is no points where the ``velocity'' $\dot D=0$, or equivalently 
there is  no points at which the 
potential term equals
the total energy. Similarly, 
using the constraint equation (3.14), one may see the scale factor present the same behavior. Therefore, in contrast with Ref.\cite{KMME}, the flat decrumpling universe 
in this approach  does not have any turning points.  
If the universe starts from a large dimension (which coresponds to a small
radius), while it expands indefinitely its dimension
decreases, however, since there is no turning point(s), the dimension
approachs continuously to its singular value $D=1$.    
 
\section{Conclusion}

   We have shown by an explicit example that the 
use of the space components of the 
Einstein field equation plus the energy conservation
law has, in principle, less information than 
the full set of Einstein's field equations. In this way, if the constraint equation has not been considered, the two calculations schemes lead, in general, to completely different physical properties. In the original decrumpling 
model, the result 
is a nonsingular and oscillatory universe,
whereas in the approach followed here the 
universe expands and its dimension decreases continuously even to less than 
$3$, approaching  
asymptotically to $D=1$ when the scale factor is infinitely large. 

In summ, the decrumpling universe worked out here is singular and evolves either in size or in the number of dimensions with no turning points.

{\bf Acknowledgment}

We are grateful to Mohammad Khorrami and 
Martin Rainer for their useful comments. This work is partially supported 
by the US Department of Energy under grant 
DE-F602-91ER40688, Task A.
One of us (JASL) is also grateful to Conselho 
Nacional de Desenvolvimento Cient\'{i}fico e Tecnol\'{o}gico - CNPq (Brazilian
Research Agency) for financial support.


\begin{thebibliography}{99}

\bibitem{QS} Quantum Structure of Space and Time,
eds. M Duff and C Isham, Cambridge University
Press, Cambridge (1982).

\bibitem{FS} V. J. Martinez, ``Fractal Aspects of 
Galaxies Clustering'', in
Applying Fractals in Astronomy, eds. A. Heck and J. M. Perdang,
Springer-Verlag, Berlim (1991).

\bibitem{CP} P. H. Coleman and L. Pietronero, Phys. Rep. 
{\bf 231}, 311 (1992); X. Luo and D. N. Schramm, Science, {\bf 256},
513 (1992). 

\bibitem{KMME} M. Khorrami, R. Mansouri, M. Mohazzab 
and M. R. Ejtehadi, ``A model Universe
with Variable Dimension: Expansion as Decrumpling'',  
gr-qc/9507059, submitted for publication.

\bibitem{BMR} U. Bleyer, M. Mohazzab, M. Rainer," Dynamics of
Dimensionin Factor Space Cosmology", GR-QC/9508035, 
 Astron.  Nachr. {\bf 317}, 3-13 (1996). 

\end{thebibliography}
\end{document}